\documentclass[twocolumn,english,aps,prl,10pt,superscriptaddress,floatfix]{revtex4-2}
\usepackage{times}
\usepackage{bm}
\usepackage{graphicx}
\usepackage{amssymb,amsfonts,amsmath,amsbsy,bm,t1enc,latexsym}
\usepackage{float}
\usepackage[colorlinks=true,citecolor=blue,linkcolor=magenta]{hyperref}
\usepackage[markup=nocolor, authormarkupposition=left]{changes}
\usepackage[english]{babel}
\usepackage{url}
\usepackage{siunitx}
\usepackage{soul}
\usepackage{changes}
\usepackage{tabularx}

\usepackage{array}
\usepackage{booktabs}

\usepackage{braket}

\begin{document}

\title{Continuous-variable quantum tomography of high-amplitude states}

\author{Ekaterina Fedotova}
\email[]{fea.fedotova@gmail.com}
\affiliation{Institute of Science and Technology Austria, am Campus 1, 3400 Klosterneuburg, Austria}

\author{Nikolai Kuznetsov}
\thanks{Current affiliation}
\affiliation{Institute of Physics, Swiss Federal Institute of Technology Lausanne (EPFL), CH-1015 Lausanne, Switzerland}

\author{Egor Tiunov}
\affiliation{Quantum Research Center, Technology Innovation Institute, Abu Dhabi, UAE}

\author{A. E. Ulanov}
 \thanks{Current affiliation}
\affiliation{Deutsches Elektronen-Synchrotron DESY, Notkestr. 85, 22607 Hamburg, Germany}

\author{A. I. Lvovsky}
\email[]{alex.lvovsky@physics.ox.ac.uk}
\affiliation{Clarendon Laboratory, University of Oxford, Parks Road, Oxford OX1 3PU, UK}

\begin{abstract}
Quantum state tomography is an essential component of modern quantum technology. In application to continuous-variable harmonic-oscillator systems, such as the electromagnetic field, existing tomography methods typically reconstruct the state in discrete bases, and are hence limited to states with relatively low amplitudes and energies. Here we overcome this limitation by utilizing a feed-forward neural network to obtain the density matrix directly in the continuous position basis. An important benefit of our approach is the ability to choose specific regions in the phase space for detailed reconstruction. This results in relatively slow scaling of the amount of resources required for the reconstruction with the state amplitude, and hence allows us to dramatically increase the range of amplitudes accessible with our method.
\end{abstract}

\maketitle



\section{Introduction}
\label{sec:Introduction}
The ongoing rapid development of quantum technology results in increased complexity of various quantum states we operate with~\cite{eaton2022measurement,nehra2021all, ourjoumtsev2006generating,eaton2019non,dakna1998quantum,thekkadath2020engineering,takase2021generation,asavanant2021wave,asavanant2017generation,asavanant2019generation,larsen2019deterministic}. This, in turn, raises the requirements to quantum state tomography (QST) --- the technique for reconstruction of the quantum state of a system from measurements~\cite{d2003quantum, d2001quantum, banaszek2013focus, lvovsky2009continuous, smithey1993measurement}. The higher dimension of the Hilbert space can exponentially enlarge the amount of data required for a QST quorum, as well as the amount of processing power required to restore the state. These problems are often solved using machine learning and neural networks, which allow QST to be carried out quickly and efficiently~\cite{ahmed2021quantum,tiunov2020experimental,torlai2018neural,carrasquilla2019reconstructing,kurmapu2022reconstructing,zhu2022flexible}.

A convenient choice of state representation for the QST with neural networks is a basis of discrete variables~\cite{torlai2018neural, ahmed2021quantum}. A discrete basis is often suitable for systems naturally described in continuous-variable (CV) bases, such as the harmonic oscillator. Indeed, for a long time, QST of harmonic oscillator states in the optical, microwave and mechanical domains has been done in the Fock basis~\cite{lvovsky2004iterative,tiunov2020experimental}, even when the measurements were performed in the continuous quadrature basis. We note that QST of harmonic oscillators historically began in the CV basis~\cite{leonhardt1997measuring} with the filtered back-projection algorithm; however, this approach  later became unpopular because of its relative computational complexity and unphysical artefacts in the reconstructed state~\cite{lvovsky2004iterative}.

However, the choice of the Fock basis for restoring the states of a harmonic oscillator is optimal only when the state under consideration has a relatively small amplitude. This is because the number of Fock terms needed to represent a state grows quadratically with its amplitude. Moreover, the superexponential factor of $(2^n n!)^{-1/2}$ present in the $n$-photon Fock state wavefunction, which enters the QST algorithm, complicates the calculations for high photon numbers. This is why the reconstruction space in the currently published results on QST of a harmonic oscillator in the Fock basis is limited to the subspace of 30 photons~\cite{tiunov2020experimental}, up to our knowledge.

In this paper, we develop a new approach to QST of a harmonic oscillator with a neural network where both the measurement and the reconstruction are in the CV domain. Specifically, we reconstruct a density matrix $ \rho(X, X') $ in the position basis on a predefined finite coordinate grid. This enables us to directly obtain density matrices of states with arbitrarily high amplitudes, overcoming computational issues. Our method allows us to on those areas of the phase space that are relevant to the state in question. For example, the wave function of a coherent state $ \ket\alpha $ takes nonzero values in a narrow region around $X_\alpha=\alpha\sqrt2$ and we can query the network with the position values mainly around $X_\alpha$, thereby ensuring that it is aware of the density matrix structure around this region particularly well. No \emph{a priori} information about the state is needed since localized regions of interest can be inferred directly from the measurement data. 
After the training, the neural network will correctly interpolate the values of the density matrix corresponding to the coordinates between the grid nodes \cite{cybenko1989approximation, hornik1991approximation}.

We anticipate our approach to be particularly useful for optical analogues of Schr\"odinger's cats, i.e.~superpositions of coherent states of different amplitudes and/or phases, which find broad application in CV quantum information processing~\cite{thekkadath2020engineering,ofek2016extending,lvovsky2020production}. Similarly to coherent states, cat states are well-localized in the quadrature space, making them amenable to our method. We demonstrate our method to perform QST of cat states with amplitudes up to $\alpha=40$, which in the Fock basis would require a reconstruction space with up to $\sim1800$ photons.

\begin{figure*}[!ht]
\includegraphics[width=1\linewidth]
{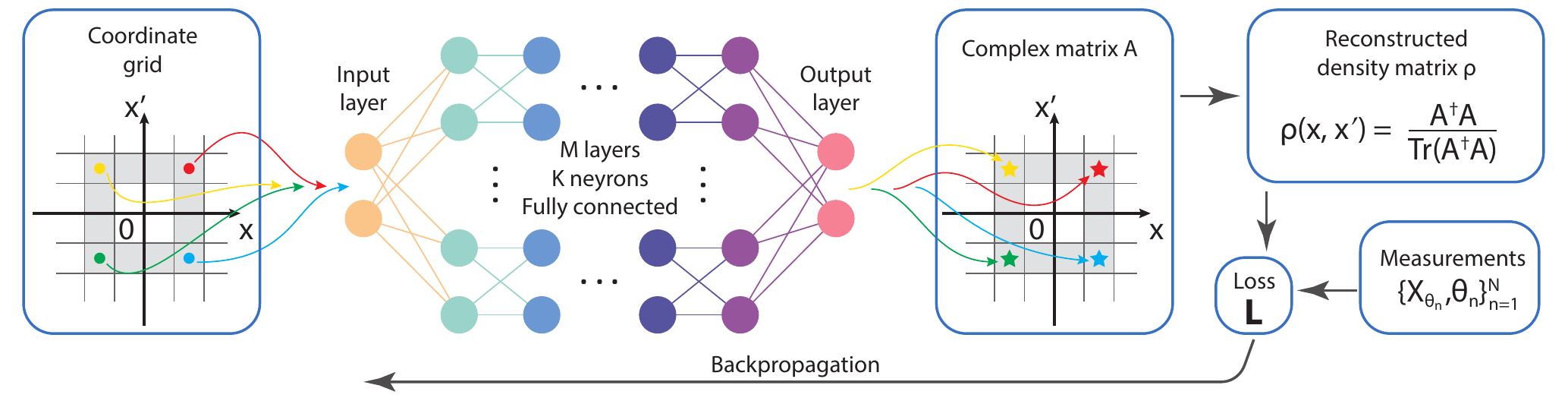}
\caption{CV QST via a fully connected feed-forward neural network. Input position value pairs $(X,X')$ within the regions of interests (shaded grey) are processed with the NN into real and imaginary values of a matrix $A(X,X')$, which is then transformed into a physically valid density matrix via the Cholesky decomposition \eqref{Cholesky}. The NN is trained using the backpropagation algorithm to maximize the log-likelihood of the measured quadrature set.}
\label{fig:NN_scheme}
\end{figure*}




\section{Concept}
\label{sec:QuantumTomography}
We specialize to optical homodyne tomography~\cite{smithey1993measurement, lvovsky2009continuous}, which measures samples of generalized quadrature~$\hat X_{\theta} =\hat{X} \cos \theta + \hat{P} \sin \theta$, where $\theta$ is the phase of the local oscillator and can be controlled in the experiment (our treatment is straightforwardly generalized to CV-QST based on other types of measurement). $N$ homodyne measurements produce the set of amplitude and phase pairs~$\set{ X_{\theta_{n}}, \theta_{n}}_{n=1}^{N} $. 
We utilise likelihood maximization approach with the log-likelihood functional which in our case is defined as
\begin{equation}\label{L}
    \mathbf{L} = - \sum_{n} \ln \left[ P(X_{\theta_{n}}, \theta_n) \right],
\end{equation}
where $ P(X_{\theta}, \theta) $ is the probability density of obtaining a measurement result $X_{\theta}$ for the local oscillator phase equal to $\theta$. Minimization of $ \mathbf{L} $ corresponds to finding the density matrix that maximizes the probability to get the specified quadrature distribution. This probability density is given (see Refs.~\cite{man1999diffraction} and Appendix) by 
\begin{subequations}
\begin{align}\label{eq:quadr_distrib_x}
    P(X_{\theta}, \theta) &= \frac{1}{2 \pi |\sin \theta|}  \iint   \rho (x, x')\cdot\\&\cdot \exp \left[ -i \dfrac{x - x'}{\sin \theta} \left( X_{\theta} -  \cos \theta \dfrac{x + x'}{2}  \right) \right]dx dx'.\nonumber
\end{align}
where  the density matrix $\rho(x,x')$ in the position basis is known from the neural network output. 

The above equation can give rise to numerical instabilities for small $\theta$ due to the denominator containing $\sin \theta$. To avoid the problem, we use Eq.~\eqref{eq:quadr_distrib_x} only for the quadratures with $|\sin \theta| \geq 1/\sqrt{2}$. When  $|\sin \theta| < 1/\sqrt{2}$, we first apply the Fourier transform to compute the density matrix $\rho (p, p')$ in the momentum basis, and then find the probabilities according to
\begin{align}\label{eq:quadr_distrib_p}
    P(X_{\theta}, \theta)  &= \dfrac{1}{2 \pi |\cos \theta|}   \iint  \rho (p, p')\cdot \\ &\cdot  \exp \left[ -i \dfrac{p - p'}{\cos \theta} \left( X_\theta +  \sin \theta \dfrac{p + p'}{2}  \right) \right] \text{d}p \text{d}p'.\nonumber
\end{align}
\end{subequations}


\begin{figure*}[!ht]
\center{\includegraphics[width=1\linewidth]{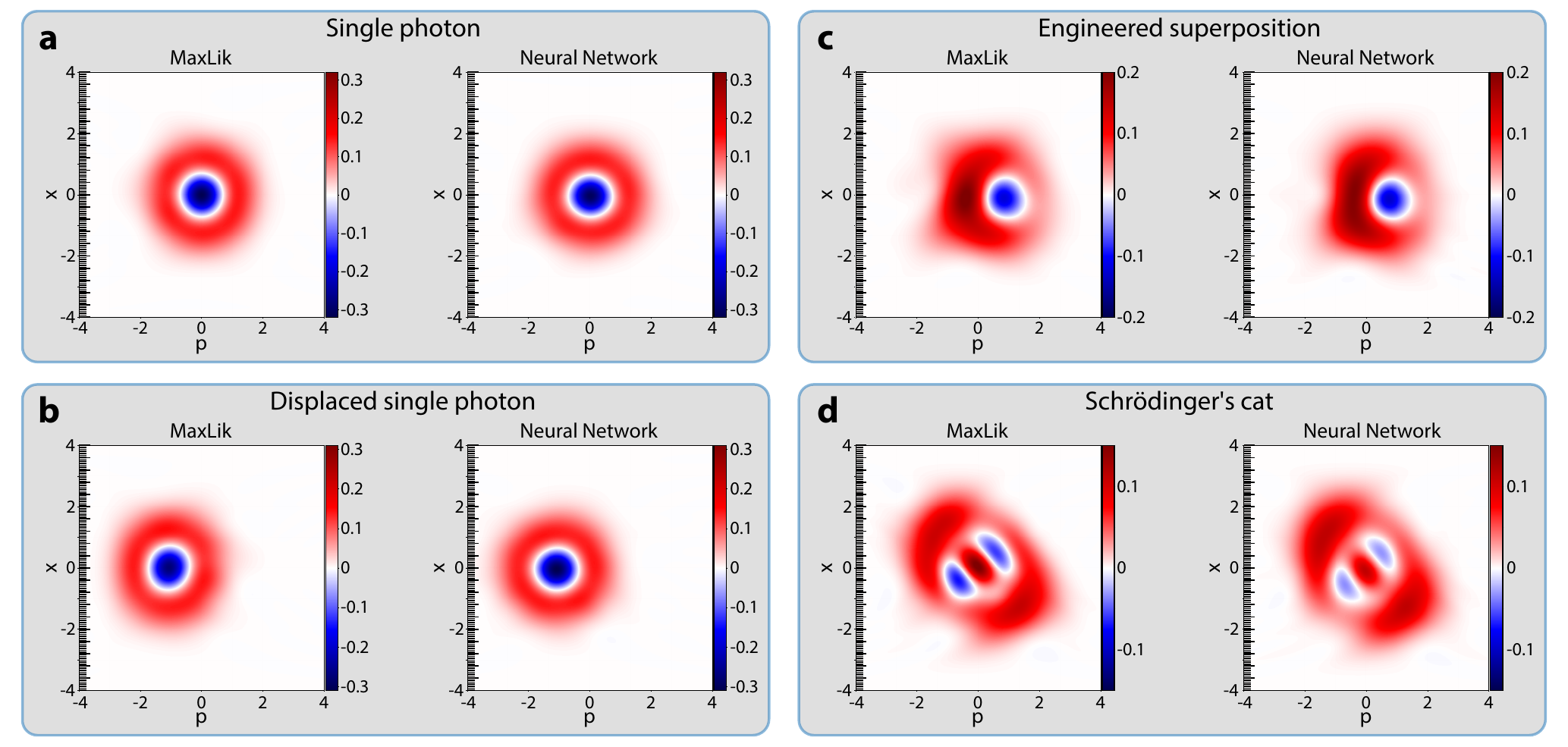}}
\caption{Wigner functions reconstructed from the experimental data with the MaxLik method and our neural network for a single photon state \textbf{(a)}, displaced single photon state \textbf{(b)}, engineered Fock superpositions up to the two-photon level \textbf{(c)} and optical Schr\"odinger’s cat \textbf{(d)}. Efficiency correction is applied to each state (see Table I). Mutual fidelity between each pair of states exceeds $ 0.994 $.}
\label{fig:exp_res}
\end{figure*}


To reconstruct the density matrix, we exploit a fully connected feed-forward neural network (NN)~\cite{svozil1997introduction} based on freely accessible \mbox{PyTorch} libraries. The  reconstruction process is shown in Fig.~\ref{fig:NN_scheme}. NN takes a pair of coordinates  as input and outputs a single complex number $A(X, X')$, which is connected to the density matrix via the Cholesky decomposition \cite{higham1990analysis}
\begin{equation}\label{Cholesky}
    \rho (X, X') = \dfrac{A^{\dagger}A}{{\rm Tr}(A^{\dagger}A)}.
\end{equation} 
The motivation for this intermediate step is to ensure that the output density matrix is Hermitian, semipositive definite and normalized~\cite{blum2012density}. 

In order to compute the density matrix, the NN is applied in sequence to all pairs $\left(X, X' \right)$ from a predefined grid. When all $A(X, X')$ are known, $\rho (X, X')$ is calculated via Eq.~\eqref{Cholesky}. To train the NN, we evaluate the loss functional $ \mathbf{L} $, and iteratively apply backpropagation to update the NN parameters.

To choose the grids for the position and momentum quadratures, we inspect the experimental data for $\theta\approx0$ and $\theta\approx\pi/2$ and find the regions where the measured samples are localized. The grids must cover these regions. The grid period ($\delta X,\delta P$) is chosen to ensure correct Fourier transform. That is, large values of momentum quadratures present in the state imply that the density matrix in the position space undergoes fast  oscillations, and vice versa. 
The grid must be sufficiently frequent to capture these oscillations. We observed correct reconstruction by setting the grid periods according to 
\begin{align*}
    \delta X&\lesssim P_{\max}^{-1};\\
    \delta P&\lesssim X_{\max}^{-1},
\end{align*}
where $X_{\max}$ and $P_{\max}$ are the highest quadrature values observed in the measurement.

\section{Results}
\label{sec:Results}

First, we test our method on several experimentally acquired sets of measurements that correspond to prepared quantum optical states containing only a few photons. The experimental apparatus is described e.g.~in Refs.~\cite{bimbard2010quantum,sychev2017enlargement}. We compare our technique with the discrete iterative likelihood-maximization algorithm (MaxLik) \cite{lvovsky2004iterative}, which reconstructs the state in the Fock basis. We apply correction for linear losses as described in the Appendix. The reconstruction NN featured three hidden layers, each containing 100 units. For the NN training, the grid in the position and momentum spaces is chosen to cover the interval $P,X\in[-4,4]$ with 80 equidistant intervals. 

The states are listed in Table I and the results of the reconstruction are shown in Fig.~\ref{fig:exp_res}. In all cases, the mutual fidelity between the density matrices obtained with the two methods exceeds $ 0.994 $. To evaluate the fidelity more precisely, we exploit the interpolation capability of the NN to predict the density matrix values over a more frequent grid than during the training. Specifically, the grid contains 400 equidistant position values over the interval $[-4,4]$.
\begin{table}\label{statetab}
\caption{Experimental states reconstructed in Fig.~\ref{fig:exp_res}. The reconstruction involves correction for linear losses corresponding to the efficiency $\eta$. The fidelity shown is between the reconstructions via the NN and MaxLik~\cite{lvovsky2004iterative} techniques.}
\begin{tabularx}{\columnwidth}{c|p{6cm}|c|c}
\hline
Fig.& \parbox{6cm}{State}& $\eta$ & Fidelity\\
\hline
\ref{fig:exp_res}(a) & \parbox{6cm}{single-photon} & 0.56 & $>0.999$\\
\hline
\ref{fig:exp_res}(b) & \parbox{6cm}{displaced single-photon} & 0.56 & $>0.999$\\
\hline
\ref{fig:exp_res}(c) & \parbox{6cm}{normalized superposition of Fock states\\~$a_0 \ket{0} + a_1 \ket{1} + a_2 \ket{2}$ with $a_0: a_1: a_2 \sim  -0.76:0.49:0.42$ \cite{bimbard2010quantum}}& 0.56 & $0.998$\\
\hline
\ref{fig:exp_res}(d) & \parbox{6cm}{normalized superposition of coherent states\\~$\ket{\alpha} + \ket{-\alpha}$ with~$ |\alpha| = 1.85$ squeezed by~$3$~dB~\cite{sychev2017enlargement}} & 0.62 & $0.994$\\
\end{tabularx}
\end{table}

\begin{figure*}[!ht]
\begin{center} 
\includegraphics[width=1\linewidth]{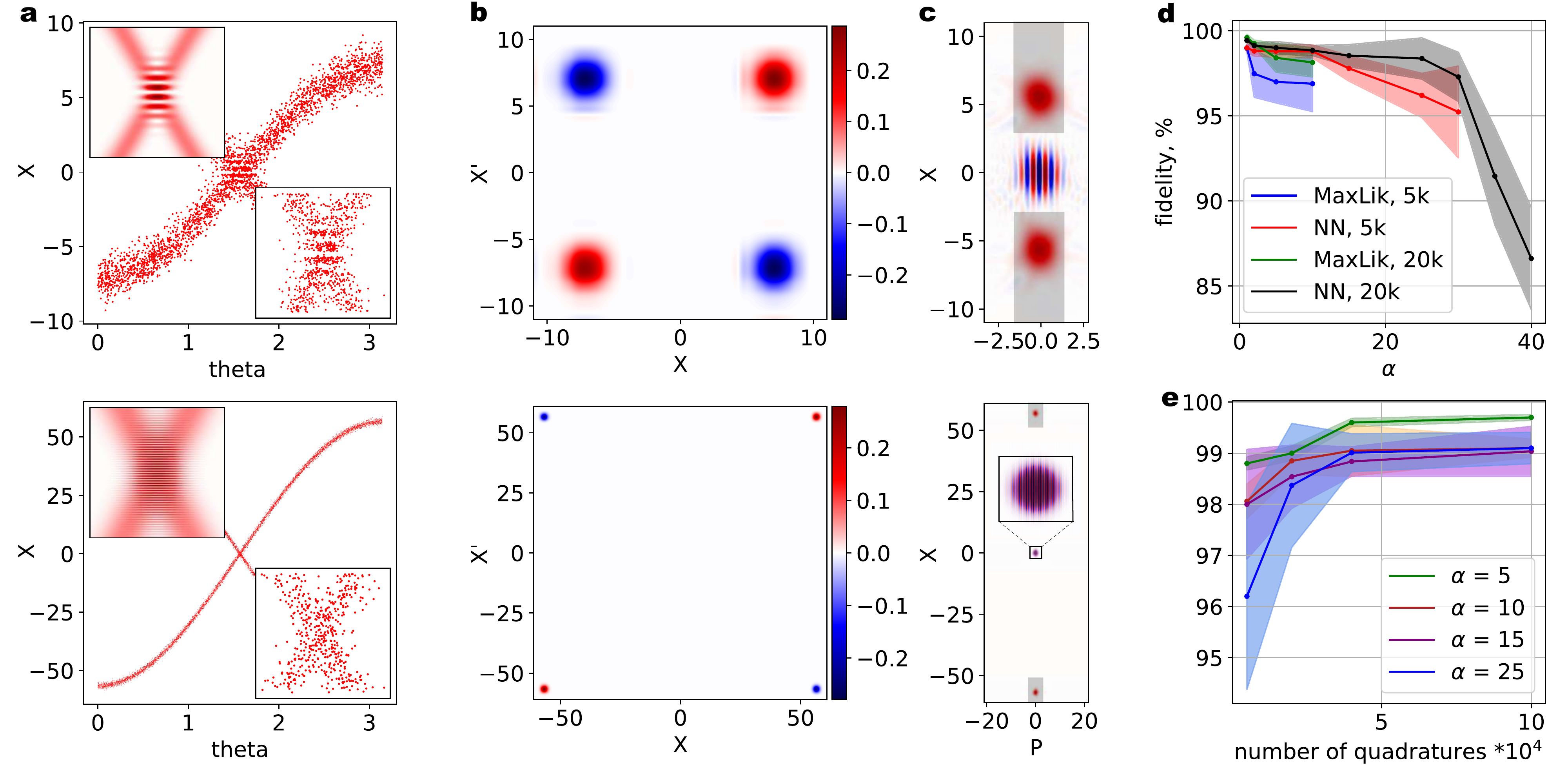}
\caption{QST of Schr\"odinger's cat states $\ket{\alpha} - \ket{-\alpha}$ with MaxLik and CN NN from simulated quadrature data sets. \textbf{(a)}: Simulated quadrature data for $\alpha = 5,40$ and $N = 5\cdot 10^3, 20\cdot 10^3$ (top and bottom panels, respectively). Bottom right insets show the zoomed-in quadrature samples in the intersection region. Top right insets present ideal probability distributions from the same regions, exhibiting fringe patterns. \textbf{(b)}: Real parts of the reconstructed density matrices for the data in \textbf{(a)}. \textbf{(c)}: Reconstructed Wigner functions for the same states. The inset in the bottom panel shows a fringe pattern in the central peak. Grey rectangles show the reconstruction regions. \textbf{d)} Fidelity as a function of the amplitude $\alpha$ for a fixed number of quadrature measurements ($ 5 \cdot 10^{3} $ and $ 20 \cdot 10^{3} $). \textbf{(e)}: dependence of the fidelity on the number of quadrature measurements for fixed amplitudes. }
\label{fig:theor_res}
\end{center}
\end{figure*}


\begin{figure*}[!ht]
\begin{center} 
\center{\includegraphics[width=1\linewidth]{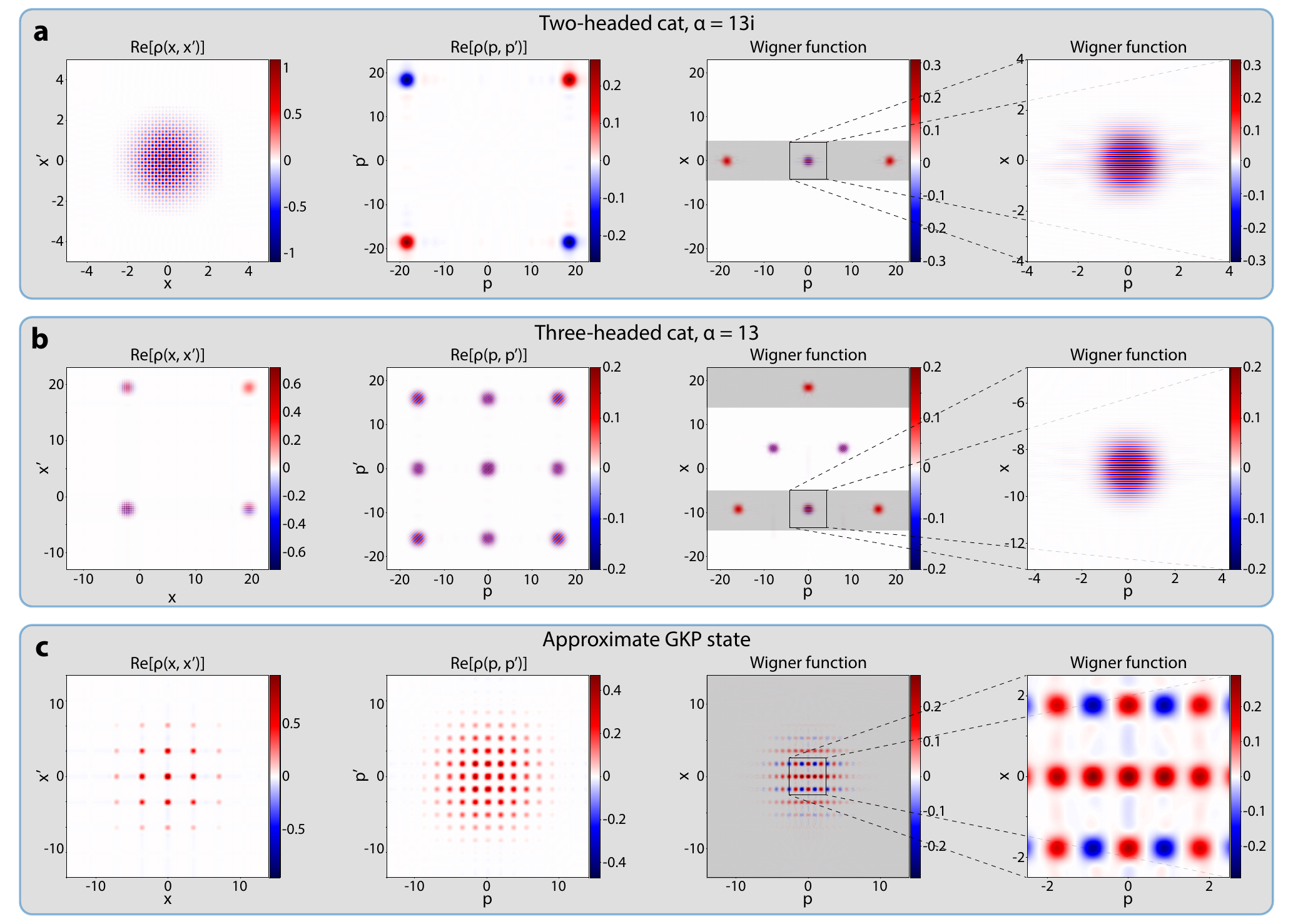}}
\caption{Various reconstructed states. \textbf{(a)}: Two-headed Schr\"odinger's cat with imaginary amplitude~$\alpha = 13i$. A checkerboard pattern is observed in the position-basis density matrix. \textbf{(b)}: Three-headed cat state with amplitude~$\alpha = 13$. The Wigner function contains three Gaussian peaks and three oscillating regions between each pair of Gaussian peaks. \textbf{(c)}: approximate Gottesman-Kitaev-Preskill state.  Grey rectangles in the Wigner function plots show the reconstruction regions. }
\label{fig:various_states}
\end{center} 
\end{figure*}

To explore our method further and demonstrate the performance of the CV NN QST approach in its full glory, we simulate quadrature measurement data sets for a variety of high-amplitude states. In Fig.~\ref{fig:theor_res} we show Schr\"odinger's cat states $\ket{\alpha} - \ket{-\alpha}$ with different real amplitudes $\alpha$. These states' Wigner functions exhibit an oscillating pattern near the phase space origin [Fig.~\ref{fig:theor_res}(c)]. The density matrix in the coordinate representation consists of two positive and two negative Gaussian peaks as shown in Fig.~\ref{fig:theor_res}(b). In the momentum basis, the density matrix exhibits a rapidly oscillating pattern with a Gaussian envelope centered around $ p = 0, p' = 0 $. 

We simulated quadrature measurement datasets with sizes ranging from $5\cdot10^3$ to $10^5$ for cat amplitudes up to $\alpha=40$. The same NN as in the previous section was used. The grid in the position space contains $ 360 $ values distributed evenly over the two intervals $[\pm\alpha\sqrt 2- 4.5,\pm\alpha\sqrt 2+ 4.5]$, i.e.~in the vicinity of the expected Gaussian peaks. In the momentum basis, the grid is in the interval $[-5,5]$, also with 360 equidistant values. 

Figures~\ref{fig:theor_res}(d,e) demonstrate the fidelity between the reconstructed and true cat states as functions of the amplitude and number of quadrature measurements. The value of each point is estimated using 5 sets of synthetically generated measurements. As expected, the fidelity increases with the number of quadratures acquired and decreases with the cat amplitude. Again, we compare the NN method with MaxLik. The latter method works with cat amplitudes up to~$\alpha = 10$. Higher amplitudes are inaccessible for MaxLik because of the aforementioned  factor of $(2^n n!)^{-1/ 2}$ in the position-basis wavefunction of a Fock state $ \ket{n} $. For instance, to represent a cat state with the amplitude~$\alpha = 10$, the reconstruction Fock basis must include elements up to at least $n = 170$, for which this numerical factor is as low as $10^{-179}$, requiring very high bit depth for precise calculation. This issue does not arise in the continuous-variable basis, making our approach advantageous for high-amplitude states. 

To illustrate the challenge of cat state reconstruction, we refer to Fig.~\ref{fig:theor_res}(a), which shows the simulated phase-dependent quadrature data. Visible fringes at the intersection of the two sinusoidal envelopes is the feature that distinguishes a cat state from an incoherent mixture of two coherent states. With growing amplitudes, the phase regions in which these intersections are present reduce in width. At the same time, the frequency of the fringes increases. This implies the requirement for larger quadrature sample sizes to make this feature statistically significant. At the same time, increasingly finer grids in the momentum space are needed, as well as high homodyne photodetection efficiencies. These factors make the required data acquisition and reconstruction process increasingly complicated and time-intensive. 

For this reason,  the magnitude of the off-diagonal peaks in the density matrix [blue peaks in Fig.~\ref{fig:theor_res}(b)], as well as the amplitude of the interference fringe pattern in the Wigner function, degrade with the growing amplitude is. It is remarkable, however, that although the fringe pattern is not visible in the dataset of Fig.~\ref{fig:theor_res}(a,bottom), the NN is able to make it out, reconstructing the cat state with a fidelity of  $0.87$ with pronounced coherence between the peaks [bottom panels in Fig.~\ref{fig:theor_res}(b,c)] in spite of a relatively small number of quadrature samples ($2\cdot 10^4$). 


Finally, we perform tomography of more complex states relevant to continuous-variable quantum information processing (Fig.~\ref{fig:various_states} ). We consider the following states.
\begin{itemize}
\item Two-headed Schr\"odinger's cats with imaginary amplitudes [(Fig.~\ref{fig:various_states}(a)]. The state is analogous to that studied above, but the position and momentum are exchanged --- so the NN needs to predict a density matrix with fast oscillations. The NN has four hidden layers with 100 units each. The activation functions are $\sin(\cdot)$ in the first two layers and $\tanh(\cdot.)$ for the second two layers. The sinusoidal activation function is helpful in modeling the oscillating behavior of the wavefunction. High-quality reconstruction demonstrates that our algorithm's capabilities are insensitive to the choice of the zero phase reference point.
\item Three-headed cat states $\ket{\alpha} + \ket{\alpha \cdot \exp{(i2\pi/3)}} + \ket{\alpha \cdot \exp{(-i2\pi/3)}}$ containing three peaks with evenly distributed phases [(Fig.~\ref{fig:various_states}(b)]. We used the same NN architecture as for the above two-headed state.
\item Gottesman-Kitaev-Preskill state ~\cite{gottesman2001encoding}, which is of importance for quantum computing~\cite{glancy2006error,glancy2009generation} [(Fig.~\ref{fig:various_states}(c)]. Ideally, the wavefunction of this state is an infinite array of delta-functions, but the approximate wavefunction relevant to experiments is an array of Gaussian peaks under a broader Gaussian envelope. The specific state used in our experiment has the wavefunction 
\begin{equation*}
    \psi_{GKP}(x) \sim \sum_{s=-\infty}^{\infty} e^{-2\pi (s k)^2 }
    e^{-\frac{(x-2s\sqrt{\pi})^2}{\Delta^2}}
\end{equation*}
where $\Delta = k = 1/4$. The NN has four hidden layers with 100 units each. The activation functions are $\sin(\cdot)$ in the third layer and $\tanh(\cdot)$ for the other layers. 
\end{itemize}
The reconstruction grid parameters and fidelities are given in Table II. As evidenced  Although some artefacts are present, our NN accurately reconstructs all of these states. 
\begin{table*}[!htbp]
\centering
\caption{Reconstruction grids and fidelities of complex state QST (Fig.~\ref{fig:various_states}). The fidelities obtained with our method ($F_{\rm NN}$) and MaxLik ($F_{\rm MaxLik}$) are shown.}
\begin{tabular}{c|c|c|c|c}
\midrule
State& position grid & momentum grid& $F_{\rm NN}$ & $F_{\rm MaxLik}$\\
\midrule
Two-headed cat, $\alpha=13i$&$[-5,5]$, 500 pts& $[-23,23]$, 500 pts& 0.98 & N/A\\
\midrule
Three-headed cat, $\alpha=13$&$[-13,-5]\cup[14,23]$, 300 pts&$[-23,23]$, 300 pts & 0.91 & N/A\\
\midrule
Gottesman-Kitaev-Preskill&$[-14,14]$, 600 pts&$[-14,14]$, 600 pts& 0.92 & 0.87\\
\bottomrule
\end{tabular}

\label{Tab:fid_comparison}
\end{table*}

\section{Conclusion}
\label{sec:Conclusion}

We demonstrate a new approach to quantum tomography based on the representation of a density matrix in a continuous-variable basis by a feed-forward neural network. The symbiosis of a continuous-variable basis and NN as universal approximator allows us to overcome the limitations on reconstructing quantum states with high amplitudes and/or photon numbers. We believe our method to be especially valuable in the context of quantum state engineering in superconducting circuits, which are known to be capable of producing high-amplitude continuous-variable states with high efficiencies \cite{vlastakis2013deterministically}.

A particularly useful feature of our NN-based method is its ability to reconstruct states in the predefined regions of the coordinate space. However, this feature cannot be taken advantage of if the state is not well-localized in the phase space. In this case, the reconstruction would require a coordinate grid that is broad and dense at the same time, and, consequently, more computing power.

Further development might be towards utilisation of more sophisticated architectures of deep learning models to represent multi-mode states and states with complex spatial structure.

\section{Appendix}
\label{sec:Methods}

\subsection{Quadrature probability distribution}
\label{subsec:MethodsQuadProbDistrib}

Here we present the alternative analytical derivation of the quadrature probability density function \eqref{eq:quadr_distrib_x}, which might be found simpler than the original derivation by Man'ko {\it et~al.}~\cite{mancini1996symplectic,man1998time,man1999diffraction}.  

First, consider the overlap between a quadrature state $\ket{X_\theta,\theta}$ and a position state $\ket{x}$. The state $\ket{X_\theta,\theta}$ is the eigenstate of the position operator in the rotated frame characterized by angle $\theta$ with eigenvalue $X_\theta$:
\begin{equation*}
    (\hat{X}\cos \theta + \hat{P}\sin \theta) \ket{X_\theta,\theta} =
    X_\theta \ket{X_\theta,\theta}.
\end{equation*}
In the position basis we can express the action of the momentum operator as $-i\dfrac{d}{dx}$, so the above equation becomes
\begin{equation*}
    \dfrac{d}{dx}\braket{x|X_\theta,\theta} = \dfrac{i}{\sin \theta}
     \left(X_\theta - x \cos \theta \right)\braket{x|X_\theta,\theta}.
\end{equation*}
The solution of this differential equation is
\begin{equation}\label{XXtheta}
    \braket{x|X_\theta,\theta} = C \exp \left( i\frac{x}{\sin \theta} X_\theta - i \tan \theta \frac{x^2}{2} \right),
\end{equation}
with $C = 1/\sqrt{2\pi |\sin \theta|}$ being a normalization factor obtained from 
\begin{equation}\braket{X'_\theta,\theta|X_\theta,\theta} = \delta(X_\theta - X'_\theta).\label{norm}
\end{equation}
Finally, the probability density $P(X_\theta,\theta)$ of observing the particular quadrature $X_\theta$ for a density matrix $\hat{\rho}(x, x')$ is
\begin{equation*}
    P(X_\theta,\theta) = \iint \braket{X_\theta,\theta|x} \rho(x,x')\braket{x'|X_\theta,\theta}dx dx'.
\end{equation*}
Substituting $\braket{x|X_\theta,\theta}$ from Eqs.~\eqref{XXtheta} and \eqref{norm}, we obtain the final result \eqref{eq:quadr_distrib_x}:
\begin{align*}
    P(X_{\theta}, \theta) &= \frac{1}{2 \pi |\sin \theta|}  \iint   \rho (x, x')\cdot\\&\cdot \exp \left[ -i \dfrac{x - x'}{\sin \theta} \left( X_{\theta} -  \cos \theta \dfrac{x + x'}{2}  \right) \right]dx dx'.
\end{align*}

\subsection{Correction for losses}
\label{subsec:LossCorr}
Here we derive the formalism tthat would enable our QST algorithm to correct for the effect of  losses and inefficient photodetection that may affect quadrature measurements. In presence of these effects, the probability distribution for the measured quadrature $X_\theta$ is~\cite{leonhardt1997measuring}
\begin{equation}
\begin{gathered}
     P(X_{\theta},\theta, \eta) = \dfrac{1}{\sqrt{\pi (1 - \eta)}} \cdot \\ \cdot \int  P(Q_{\theta},\theta) \exp \left[ - \dfrac{\eta}{1 - \eta} \left( Q_{\theta} - \dfrac{X_{\theta}}{\sqrt{\eta}} \right)^{2} \right] dQ_{\theta},
\end{gathered}
\end{equation}
where $P(Q_{\theta},\theta)$ is the quadrature probability density in the absence of losses given by Eq.~\eqref{eq:quadr_distrib_x}. After some simplifications, we obtain
\begin{equation}
\begin{gathered} \label{eq:prob_with_correction}
     P(X_{\theta},\theta, \eta) = \dfrac{1}{2\pi \sqrt{\eta} |\sin \theta|}  \iint  \rho (x, x')  \cdot \\ \cdot  \exp \left\{  - \dfrac{x - x'}{\sin \theta}  \left[ \dfrac{(x - x')(1 - \eta)}{4 \eta \sin \theta}  \right.\right.  + \\ + \left.\left. i \left(  \dfrac{X_{\theta}}{\sqrt{\eta}} - \cos \theta \dfrac{x + x'}{2}   \right) \right]  \right\} dx dx'.
\end{gathered}
\end{equation}
A similar expression can be derived for the momentum basis. These expressions are then used in place of Eqs.~\eqref{eq:quadr_distrib_x} and \eqref{eq:quadr_distrib_p} to compute the likelihood \eqref{L}. 
Remarkably, Eq.~\eqref{eq:prob_with_correction} contains integration over as many variables as the lossless integral \eqref{eq:quadr_distrib_x}. This is in contrast to the  iterative algorithm~\cite{lvovsky2004iterative}, which requires additional summation over all basis elements.







\section*{Author Contributions}
ET came up with the project; EF and NK built NNs and performed simulations; EF, AU and NK collected the data; all authors prepared the manuscript; ET and AL supervised the project.

\section*{Acknowledgments} 
The authors thank V.I. Man'ko for a fruitful discussion.

\section*{Data availability} 
The code and experimental data used for tomography within this work will be available upon reasonable request after publication of this preprint.

\section*{Disclosures}
All authors declare no conflict of interest.\\
\\

\bibliography{bibliography}

\begin{thebibliography}{40}%
\makeatletter
\providecommand \@ifxundefined [1]{%
 \@ifx{#1\undefined}
}%
\providecommand \@ifnum [1]{%
 \ifnum #1\expandafter \@firstoftwo
 \else \expandafter \@secondoftwo
 \fi
}%
\providecommand \@ifx [1]{%
 \ifx #1\expandafter \@firstoftwo
 \else \expandafter \@secondoftwo
 \fi
}%
\providecommand \natexlab [1]{#1}%
\providecommand \enquote  [1]{``#1''}%
\providecommand \bibnamefont  [1]{#1}%
\providecommand \bibfnamefont [1]{#1}%
\providecommand \citenamefont [1]{#1}%
\providecommand \href@noop [0]{\@secondoftwo}%
\providecommand \href [0]{\begingroup \@sanitize@url \@href}%
\providecommand \@href[1]{\@@startlink{#1}\@@href}%
\providecommand \@@href[1]{\endgroup#1\@@endlink}%
\providecommand \@sanitize@url [0]{\catcode `\\12\catcode `\$12\catcode
  `\&12\catcode `\#12\catcode `\^12\catcode `\_12\catcode `\%12\relax}%
\providecommand \@@startlink[1]{}%
\providecommand \@@endlink[0]{}%
\providecommand \url  [0]{\begingroup\@sanitize@url \@url }%
\providecommand \@url [1]{\endgroup\@href {#1}{\urlprefix }}%
\providecommand \urlprefix  [0]{URL }%
\providecommand \Eprint [0]{\href }%
\providecommand \doibase [0]{https://doi.org/}%
\providecommand \selectlanguage [0]{\@gobble}%
\providecommand \bibinfo  [0]{\@secondoftwo}%
\providecommand \bibfield  [0]{\@secondoftwo}%
\providecommand \translation [1]{[#1]}%
\providecommand \BibitemOpen [0]{}%
\providecommand \bibitemStop [0]{}%
\providecommand \bibitemNoStop [0]{.\EOS\space}%
\providecommand \EOS [0]{\spacefactor3000\relax}%
\providecommand \BibitemShut  [1]{\csname bibitem#1\endcsname}%
\let\auto@bib@innerbib\@empty
\bibitem [{\citenamefont {Eaton}\ \emph {et~al.}(2022)\citenamefont {Eaton},
  \citenamefont {Gonz{\'a}lez-Arciniegas}, \citenamefont {Alexander},
  \citenamefont {Menicucci},\ and\ \citenamefont
  {Pfister}}]{eaton2022measurement}%
  \BibitemOpen
  \bibfield  {author} {\bibinfo {author} {\bibfnamefont {M.}~\bibnamefont
  {Eaton}}, \bibinfo {author} {\bibfnamefont {C.}~\bibnamefont
  {Gonz{\'a}lez-Arciniegas}}, \bibinfo {author} {\bibfnamefont {R.~N.}\
  \bibnamefont {Alexander}}, \bibinfo {author} {\bibfnamefont {N.~C.}\
  \bibnamefont {Menicucci}},\ and\ \bibinfo {author} {\bibfnamefont
  {O.}~\bibnamefont {Pfister}},\ }\bibfield  {title} {\bibinfo {title}
  {Measurement-based generation and preservation of cat and grid states within
  a continuous-variable cluster state},\ }\href
  {https://quantum-journal.org/papers/q-2022-07-20-769/} {\bibfield  {journal}
  {\bibinfo  {journal} {Quantum}\ }\textbf {\bibinfo {volume} {6}},\ \bibinfo
  {pages} {769} (\bibinfo {year} {2022})}\BibitemShut {NoStop}%
\bibitem [{\citenamefont {Nehra}\ \emph {et~al.}(2021)\citenamefont {Nehra},
  \citenamefont {Eaton}, \citenamefont {Pfister},\ and\ \citenamefont
  {Marandi}}]{nehra2021all}%
  \BibitemOpen
  \bibfield  {author} {\bibinfo {author} {\bibfnamefont {R.}~\bibnamefont
  {Nehra}}, \bibinfo {author} {\bibfnamefont {M.}~\bibnamefont {Eaton}},
  \bibinfo {author} {\bibfnamefont {O.}~\bibnamefont {Pfister}},\ and\ \bibinfo
  {author} {\bibfnamefont {A.}~\bibnamefont {Marandi}},\ }\bibfield  {title}
  {\bibinfo {title} {All-optical quantum state engineering for
  rotation-symmetric bosonic states},\ }\href
  {https://arxiv.org/abs/2105.11035} {\bibfield  {journal} {\bibinfo  {journal}
  {arXiv preprint arXiv:2105.11035}\ } (\bibinfo {year} {2021})}\BibitemShut
  {NoStop}%
\bibitem [{\citenamefont {Ourjoumtsev}\ \emph {et~al.}(2006)\citenamefont
  {Ourjoumtsev}, \citenamefont {Tualle-Brouri}, \citenamefont {Laurat},\ and\
  \citenamefont {Grangier}}]{ourjoumtsev2006generating}%
  \BibitemOpen
  \bibfield  {author} {\bibinfo {author} {\bibfnamefont {A.}~\bibnamefont
  {Ourjoumtsev}}, \bibinfo {author} {\bibfnamefont {R.}~\bibnamefont
  {Tualle-Brouri}}, \bibinfo {author} {\bibfnamefont {J.}~\bibnamefont
  {Laurat}},\ and\ \bibinfo {author} {\bibfnamefont {P.}~\bibnamefont
  {Grangier}},\ }\bibfield  {title} {\bibinfo {title} {Generating optical
  schrodinger kittens for quantum information processing},\ }\href
  {https://www.science.org/doi/abs/10.1126/science.1122858} {\bibfield
  {journal} {\bibinfo  {journal} {Science}\ }\textbf {\bibinfo {volume}
  {312}},\ \bibinfo {pages} {83} (\bibinfo {year} {2006})}\BibitemShut
  {NoStop}%
\bibitem [{\citenamefont {Eaton}\ \emph {et~al.}(2019)\citenamefont {Eaton},
  \citenamefont {Nehra},\ and\ \citenamefont {Pfister}}]{eaton2019non}%
  \BibitemOpen
  \bibfield  {author} {\bibinfo {author} {\bibfnamefont {M.}~\bibnamefont
  {Eaton}}, \bibinfo {author} {\bibfnamefont {R.}~\bibnamefont {Nehra}},\ and\
  \bibinfo {author} {\bibfnamefont {O.}~\bibnamefont {Pfister}},\ }\bibfield
  {title} {\bibinfo {title} {Non-gaussian and gottesman--kitaev--preskill state
  preparation by photon catalysis},\ }\href
  {https://iopscience.iop.org/article/10.1088/1367-2630/ab5330/meta} {\bibfield
   {journal} {\bibinfo  {journal} {New Journal of Physics}\ }\textbf {\bibinfo
  {volume} {21}},\ \bibinfo {pages} {113034} (\bibinfo {year}
  {2019})}\BibitemShut {NoStop}%
\bibitem [{\citenamefont {Dakna}\ \emph {et~al.}(1998)\citenamefont {Dakna},
  \citenamefont {Kn{\"o}ll},\ and\ \citenamefont {Welsch}}]{dakna1998quantum}%
  \BibitemOpen
  \bibfield  {author} {\bibinfo {author} {\bibfnamefont {M.}~\bibnamefont
  {Dakna}}, \bibinfo {author} {\bibfnamefont {L.}~\bibnamefont {Kn{\"o}ll}},\
  and\ \bibinfo {author} {\bibfnamefont {D.-G.}\ \bibnamefont {Welsch}},\
  }\bibfield  {title} {\bibinfo {title} {Quantum state engineering using
  conditional measurement on a beam splitter},\ }\href
  {https://link.springer.com/article/10.1007/s100530050177} {\bibfield
  {journal} {\bibinfo  {journal} {The European Physical Journal D-Atomic,
  Molecular, Optical and Plasma Physics}\ }\textbf {\bibinfo {volume} {3}},\
  \bibinfo {pages} {295} (\bibinfo {year} {1998})}\BibitemShut {NoStop}%
\bibitem [{\citenamefont {Thekkadath}\ \emph {et~al.}(2020)\citenamefont
  {Thekkadath}, \citenamefont {Bell}, \citenamefont {Walmsley},\ and\
  \citenamefont {Lvovsky}}]{thekkadath2020engineering}%
  \BibitemOpen
  \bibfield  {author} {\bibinfo {author} {\bibfnamefont {G.}~\bibnamefont
  {Thekkadath}}, \bibinfo {author} {\bibfnamefont {B.}~\bibnamefont {Bell}},
  \bibinfo {author} {\bibfnamefont {I.~A.}\ \bibnamefont {Walmsley}},\ and\
  \bibinfo {author} {\bibfnamefont {A.}~\bibnamefont {Lvovsky}},\ }\bibfield
  {title} {\bibinfo {title} {Engineering schr{\"o}dinger cat states with a
  photonic even-parity detector},\ }\href
  {https://quantum-journal.org/papers/q-2020-03-02-239/} {\bibfield  {journal}
  {\bibinfo  {journal} {Quantum}\ }\textbf {\bibinfo {volume} {4}},\ \bibinfo
  {pages} {239} (\bibinfo {year} {2020})}\BibitemShut {NoStop}%
\bibitem [{\citenamefont {Takase}\ \emph {et~al.}(2021)\citenamefont {Takase},
  \citenamefont {Yoshikawa}, \citenamefont {Asavanant}, \citenamefont {Endo},\
  and\ \citenamefont {Furusawa}}]{takase2021generation}%
  \BibitemOpen
  \bibfield  {author} {\bibinfo {author} {\bibfnamefont {K.}~\bibnamefont
  {Takase}}, \bibinfo {author} {\bibfnamefont {J.-i.}\ \bibnamefont
  {Yoshikawa}}, \bibinfo {author} {\bibfnamefont {W.}~\bibnamefont
  {Asavanant}}, \bibinfo {author} {\bibfnamefont {M.}~\bibnamefont {Endo}},\
  and\ \bibinfo {author} {\bibfnamefont {A.}~\bibnamefont {Furusawa}},\
  }\bibfield  {title} {\bibinfo {title} {Generation of optical schr{\"o}dinger
  cat states by generalized photon subtraction},\ }\href
  {https://journals.aps.org/pra/abstract/10.1103/PhysRevA.103.013710}
  {\bibfield  {journal} {\bibinfo  {journal} {Physical Review A}\ }\textbf
  {\bibinfo {volume} {103}},\ \bibinfo {pages} {013710} (\bibinfo {year}
  {2021})}\BibitemShut {NoStop}%
\bibitem [{\citenamefont {Asavanant}\ \emph {et~al.}(2021)\citenamefont
  {Asavanant}, \citenamefont {Takase}, \citenamefont {Fukui}, \citenamefont
  {Endo}, \citenamefont {Yoshikawa},\ and\ \citenamefont
  {Furusawa}}]{asavanant2021wave}%
  \BibitemOpen
  \bibfield  {author} {\bibinfo {author} {\bibfnamefont {W.}~\bibnamefont
  {Asavanant}}, \bibinfo {author} {\bibfnamefont {K.}~\bibnamefont {Takase}},
  \bibinfo {author} {\bibfnamefont {K.}~\bibnamefont {Fukui}}, \bibinfo
  {author} {\bibfnamefont {M.}~\bibnamefont {Endo}}, \bibinfo {author}
  {\bibfnamefont {J.-i.}\ \bibnamefont {Yoshikawa}},\ and\ \bibinfo {author}
  {\bibfnamefont {A.}~\bibnamefont {Furusawa}},\ }\bibfield  {title} {\bibinfo
  {title} {Wave-function engineering via conditional quantum teleportation with
  a non-gaussian entanglement resource},\ }\href
  {https://journals.aps.org/pra/abstract/10.1103/PhysRevA.103.043701}
  {\bibfield  {journal} {\bibinfo  {journal} {Physical Review A}\ }\textbf
  {\bibinfo {volume} {103}},\ \bibinfo {pages} {043701} (\bibinfo {year}
  {2021})}\BibitemShut {NoStop}%
\bibitem [{\citenamefont {Asavanant}\ \emph {et~al.}(2017)\citenamefont
  {Asavanant}, \citenamefont {Nakashima}, \citenamefont {Shiozawa},
  \citenamefont {Yoshikawa},\ and\ \citenamefont
  {Furusawa}}]{asavanant2017generation}%
  \BibitemOpen
  \bibfield  {author} {\bibinfo {author} {\bibfnamefont {W.}~\bibnamefont
  {Asavanant}}, \bibinfo {author} {\bibfnamefont {K.}~\bibnamefont
  {Nakashima}}, \bibinfo {author} {\bibfnamefont {Y.}~\bibnamefont {Shiozawa}},
  \bibinfo {author} {\bibfnamefont {J.-I.}\ \bibnamefont {Yoshikawa}},\ and\
  \bibinfo {author} {\bibfnamefont {A.}~\bibnamefont {Furusawa}},\ }\bibfield
  {title} {\bibinfo {title} {Generation of highly pure schr{\"o}dinger’s cat
  states and real-time quadrature measurements via optical filtering},\ }\href
  {https://opg.optica.org/oe/fulltext.cfm?uri=oe-25-26-32227&id=379434}
  {\bibfield  {journal} {\bibinfo  {journal} {Optics Express}\ }\textbf
  {\bibinfo {volume} {25}},\ \bibinfo {pages} {32227} (\bibinfo {year}
  {2017})}\BibitemShut {NoStop}%
\bibitem [{\citenamefont {Asavanant}\ \emph {et~al.}(2019)\citenamefont
  {Asavanant}, \citenamefont {Shiozawa}, \citenamefont {Yokoyama},
  \citenamefont {Charoensombutamon}, \citenamefont {Emura}, \citenamefont
  {Alexander}, \citenamefont {Takeda}, \citenamefont {Yoshikawa}, \citenamefont
  {Menicucci}, \citenamefont {Yonezawa} \emph
  {et~al.}}]{asavanant2019generation}%
  \BibitemOpen
  \bibfield  {author} {\bibinfo {author} {\bibfnamefont {W.}~\bibnamefont
  {Asavanant}}, \bibinfo {author} {\bibfnamefont {Y.}~\bibnamefont {Shiozawa}},
  \bibinfo {author} {\bibfnamefont {S.}~\bibnamefont {Yokoyama}}, \bibinfo
  {author} {\bibfnamefont {B.}~\bibnamefont {Charoensombutamon}}, \bibinfo
  {author} {\bibfnamefont {H.}~\bibnamefont {Emura}}, \bibinfo {author}
  {\bibfnamefont {R.~N.}\ \bibnamefont {Alexander}}, \bibinfo {author}
  {\bibfnamefont {S.}~\bibnamefont {Takeda}}, \bibinfo {author} {\bibfnamefont
  {J.-i.}\ \bibnamefont {Yoshikawa}}, \bibinfo {author} {\bibfnamefont {N.~C.}\
  \bibnamefont {Menicucci}}, \bibinfo {author} {\bibfnamefont {H.}~\bibnamefont
  {Yonezawa}}, \emph {et~al.},\ }\bibfield  {title} {\bibinfo {title}
  {Generation of time-domain-multiplexed two-dimensional cluster state},\
  }\href {https://www.science.org/doi/abs/10.1126/science.aay2645} {\bibfield
  {journal} {\bibinfo  {journal} {Science}\ }\textbf {\bibinfo {volume}
  {366}},\ \bibinfo {pages} {373} (\bibinfo {year} {2019})}\BibitemShut
  {NoStop}%
\bibitem [{\citenamefont {Larsen}\ \emph {et~al.}(2019)\citenamefont {Larsen},
  \citenamefont {Guo}, \citenamefont {Breum}, \citenamefont
  {Neergaard-Nielsen},\ and\ \citenamefont
  {Andersen}}]{larsen2019deterministic}%
  \BibitemOpen
  \bibfield  {author} {\bibinfo {author} {\bibfnamefont {M.~V.}\ \bibnamefont
  {Larsen}}, \bibinfo {author} {\bibfnamefont {X.}~\bibnamefont {Guo}},
  \bibinfo {author} {\bibfnamefont {C.~R.}\ \bibnamefont {Breum}}, \bibinfo
  {author} {\bibfnamefont {J.~S.}\ \bibnamefont {Neergaard-Nielsen}},\ and\
  \bibinfo {author} {\bibfnamefont {U.~L.}\ \bibnamefont {Andersen}},\
  }\bibfield  {title} {\bibinfo {title} {Deterministic generation of a
  two-dimensional cluster state},\ }\href
  {https://www.science.org/doi/abs/10.1126/science.aay4354} {\bibfield
  {journal} {\bibinfo  {journal} {Science}\ }\textbf {\bibinfo {volume}
  {366}},\ \bibinfo {pages} {369} (\bibinfo {year} {2019})}\BibitemShut
  {NoStop}%
\bibitem [{\citenamefont {Paris}\ and\ \citenamefont
  {Sacchi}(2003)}]{d2003quantum}%
  \BibitemOpen
  \bibfield  {author} {\bibinfo {author} {\bibfnamefont {M.~G.}\ \bibnamefont
  {Paris}}\ and\ \bibinfo {author} {\bibfnamefont {M.~F.}\ \bibnamefont
  {Sacchi}},\ }\bibfield  {title} {\bibinfo {title} {Quantum tomography},\
  }\href {https://arxiv.org/abs/quant-ph/0302028} {\bibfield  {journal}
  {\bibinfo  {journal} {Advances in Imaging and Electron Physics}\ }\textbf
  {\bibinfo {volume} {128}},\ \bibinfo {pages} {206} (\bibinfo {year}
  {2003})}\BibitemShut {NoStop}%
\bibitem [{\citenamefont {D'Ariano}\ and\ \citenamefont
  {Presti}(2001)}]{d2001quantum}%
  \BibitemOpen
  \bibfield  {author} {\bibinfo {author} {\bibfnamefont {G.}~\bibnamefont
  {D'Ariano}}\ and\ \bibinfo {author} {\bibfnamefont {P.~L.}\ \bibnamefont
  {Presti}},\ }\bibfield  {title} {\bibinfo {title} {Quantum tomography for
  measuring experimentally the matrix elements of an arbitrary quantum
  operation},\ }\href
  {https://journals.aps.org/prl/abstract/10.1103/PhysRevLett.86.4195}
  {\bibfield  {journal} {\bibinfo  {journal} {Physical review letters}\
  }\textbf {\bibinfo {volume} {86}},\ \bibinfo {pages} {4195} (\bibinfo {year}
  {2001})}\BibitemShut {NoStop}%
\bibitem [{\citenamefont {Banaszek}\ \emph {et~al.}(2013)\citenamefont
  {Banaszek}, \citenamefont {Cramer},\ and\ \citenamefont
  {Gross}}]{banaszek2013focus}%
  \BibitemOpen
  \bibfield  {author} {\bibinfo {author} {\bibfnamefont {K.}~\bibnamefont
  {Banaszek}}, \bibinfo {author} {\bibfnamefont {M.}~\bibnamefont {Cramer}},\
  and\ \bibinfo {author} {\bibfnamefont {D.}~\bibnamefont {Gross}},\ }\bibfield
   {title} {\bibinfo {title} {Focus on quantum tomography},\ }\href
  {https://iopscience.iop.org/article/10.1088/1367-2630/15/12/125020/meta}
  {\bibfield  {journal} {\bibinfo  {journal} {New Journal of Physics}\ }\textbf
  {\bibinfo {volume} {15}},\ \bibinfo {pages} {125020} (\bibinfo {year}
  {2013})}\BibitemShut {NoStop}%
\bibitem [{\citenamefont {Lvovsky}\ and\ \citenamefont
  {Raymer}(2009)}]{lvovsky2009continuous}%
  \BibitemOpen
  \bibfield  {author} {\bibinfo {author} {\bibfnamefont {A.~I.}\ \bibnamefont
  {Lvovsky}}\ and\ \bibinfo {author} {\bibfnamefont {M.~G.}\ \bibnamefont
  {Raymer}},\ }\bibfield  {title} {\bibinfo {title} {Continuous-variable
  optical quantum-state tomography},\ }\href
  {https://journals.aps.org/rmp/abstract/10.1103/RevModPhys.81.299} {\bibfield
  {journal} {\bibinfo  {journal} {Reviews of modern physics}\ }\textbf
  {\bibinfo {volume} {81}},\ \bibinfo {pages} {299} (\bibinfo {year}
  {2009})}\BibitemShut {NoStop}%
\bibitem [{\citenamefont {Smithey}\ \emph {et~al.}(1993)\citenamefont
  {Smithey}, \citenamefont {Beck}, \citenamefont {Raymer},\ and\ \citenamefont
  {Faridani}}]{smithey1993measurement}%
  \BibitemOpen
  \bibfield  {author} {\bibinfo {author} {\bibfnamefont {D.}~\bibnamefont
  {Smithey}}, \bibinfo {author} {\bibfnamefont {M.}~\bibnamefont {Beck}},
  \bibinfo {author} {\bibfnamefont {M.~G.}\ \bibnamefont {Raymer}},\ and\
  \bibinfo {author} {\bibfnamefont {A.}~\bibnamefont {Faridani}},\ }\bibfield
  {title} {\bibinfo {title} {Measurement of the wigner distribution and the
  density matrix of a light mode using optical homodyne tomography: Application
  to squeezed states and the vacuum},\ }\href
  {https://journals.aps.org/prl/abstract/10.1103/PhysRevLett.70.1244}
  {\bibfield  {journal} {\bibinfo  {journal} {Physical review letters}\
  }\textbf {\bibinfo {volume} {70}},\ \bibinfo {pages} {1244} (\bibinfo {year}
  {1993})}\BibitemShut {NoStop}%
\bibitem [{\citenamefont {Ahmed}\ \emph {et~al.}(2021)\citenamefont {Ahmed},
  \citenamefont {Mu{\~n}oz}, \citenamefont {Nori},\ and\ \citenamefont
  {Kockum}}]{ahmed2021quantum}%
  \BibitemOpen
  \bibfield  {author} {\bibinfo {author} {\bibfnamefont {S.}~\bibnamefont
  {Ahmed}}, \bibinfo {author} {\bibfnamefont {C.~S.}\ \bibnamefont
  {Mu{\~n}oz}}, \bibinfo {author} {\bibfnamefont {F.}~\bibnamefont {Nori}},\
  and\ \bibinfo {author} {\bibfnamefont {A.~F.}\ \bibnamefont {Kockum}},\
  }\bibfield  {title} {\bibinfo {title} {Quantum state tomography with
  conditional generative adversarial networks},\ }\href
  {https://journals.aps.org/prl/abstract/10.1103/PhysRevLett.127.140502}
  {\bibfield  {journal} {\bibinfo  {journal} {Physical Review Letters}\
  }\textbf {\bibinfo {volume} {127}},\ \bibinfo {pages} {140502} (\bibinfo
  {year} {2021})}\BibitemShut {NoStop}%
\bibitem [{\citenamefont {Tiunov}\ \emph {et~al.}(2020)\citenamefont {Tiunov},
  \citenamefont {Tiunova}, \citenamefont {Ulanov}, \citenamefont {Lvovsky},\
  and\ \citenamefont {Fedorov}}]{tiunov2020experimental}%
  \BibitemOpen
  \bibfield  {author} {\bibinfo {author} {\bibfnamefont {E.~S.}\ \bibnamefont
  {Tiunov}}, \bibinfo {author} {\bibfnamefont {V.}~\bibnamefont {Tiunova}},
  \bibinfo {author} {\bibfnamefont {A.~E.}\ \bibnamefont {Ulanov}}, \bibinfo
  {author} {\bibfnamefont {A.}~\bibnamefont {Lvovsky}},\ and\ \bibinfo {author}
  {\bibfnamefont {A.~K.}\ \bibnamefont {Fedorov}},\ }\bibfield  {title}
  {\bibinfo {title} {Experimental quantum homodyne tomography via machine
  learning},\ }\href
  {https://opg.optica.org/optica/fulltext.cfm?uri=optica-7-5-448&id=431506}
  {\bibfield  {journal} {\bibinfo  {journal} {Optica}\ }\textbf {\bibinfo
  {volume} {7}},\ \bibinfo {pages} {448} (\bibinfo {year} {2020})}\BibitemShut
  {NoStop}%
\bibitem [{\citenamefont {Torlai}\ \emph {et~al.}(2018)\citenamefont {Torlai},
  \citenamefont {Mazzola}, \citenamefont {Carrasquilla}, \citenamefont
  {Troyer}, \citenamefont {Melko},\ and\ \citenamefont
  {Carleo}}]{torlai2018neural}%
  \BibitemOpen
  \bibfield  {author} {\bibinfo {author} {\bibfnamefont {G.}~\bibnamefont
  {Torlai}}, \bibinfo {author} {\bibfnamefont {G.}~\bibnamefont {Mazzola}},
  \bibinfo {author} {\bibfnamefont {J.}~\bibnamefont {Carrasquilla}}, \bibinfo
  {author} {\bibfnamefont {M.}~\bibnamefont {Troyer}}, \bibinfo {author}
  {\bibfnamefont {R.}~\bibnamefont {Melko}},\ and\ \bibinfo {author}
  {\bibfnamefont {G.}~\bibnamefont {Carleo}},\ }\bibfield  {title} {\bibinfo
  {title} {Neural-network quantum state tomography},\ }\href
  {https://www.nature.com/articles/s41567-018-0048-5} {\bibfield  {journal}
  {\bibinfo  {journal} {Nature Physics}\ }\textbf {\bibinfo {volume} {14}},\
  \bibinfo {pages} {447} (\bibinfo {year} {2018})}\BibitemShut {NoStop}%
\bibitem [{\citenamefont {Carrasquilla}\ \emph {et~al.}(2019)\citenamefont
  {Carrasquilla}, \citenamefont {Torlai}, \citenamefont {Melko},\ and\
  \citenamefont {Aolita}}]{carrasquilla2019reconstructing}%
  \BibitemOpen
  \bibfield  {author} {\bibinfo {author} {\bibfnamefont {J.}~\bibnamefont
  {Carrasquilla}}, \bibinfo {author} {\bibfnamefont {G.}~\bibnamefont
  {Torlai}}, \bibinfo {author} {\bibfnamefont {R.~G.}\ \bibnamefont {Melko}},\
  and\ \bibinfo {author} {\bibfnamefont {L.}~\bibnamefont {Aolita}},\
  }\bibfield  {title} {\bibinfo {title} {Reconstructing quantum states with
  generative models},\ }\href
  {https://www.nature.com/articles/s42256-019-0028-1} {\bibfield  {journal}
  {\bibinfo  {journal} {Nature Machine Intelligence}\ }\textbf {\bibinfo
  {volume} {1}},\ \bibinfo {pages} {155} (\bibinfo {year} {2019})}\BibitemShut
  {NoStop}%
\bibitem [{\citenamefont {Kurmapu}\ \emph {et~al.}(2022)\citenamefont
  {Kurmapu}, \citenamefont {Tiunova}, \citenamefont {Tiunov}, \citenamefont
  {Ringbauer}, \citenamefont {Maier}, \citenamefont {Blatt}, \citenamefont
  {Monz}, \citenamefont {Fedorov},\ and\ \citenamefont
  {Lvovsky}}]{kurmapu2022reconstructing}%
  \BibitemOpen
  \bibfield  {author} {\bibinfo {author} {\bibfnamefont {M.~K.}\ \bibnamefont
  {Kurmapu}}, \bibinfo {author} {\bibfnamefont {V.}~\bibnamefont {Tiunova}},
  \bibinfo {author} {\bibfnamefont {E.}~\bibnamefont {Tiunov}}, \bibinfo
  {author} {\bibfnamefont {M.}~\bibnamefont {Ringbauer}}, \bibinfo {author}
  {\bibfnamefont {C.}~\bibnamefont {Maier}}, \bibinfo {author} {\bibfnamefont
  {R.}~\bibnamefont {Blatt}}, \bibinfo {author} {\bibfnamefont
  {T.}~\bibnamefont {Monz}}, \bibinfo {author} {\bibfnamefont {A.~K.}\
  \bibnamefont {Fedorov}},\ and\ \bibinfo {author} {\bibfnamefont
  {A.}~\bibnamefont {Lvovsky}},\ }\bibfield  {title} {\bibinfo {title}
  {Reconstructing complex states of a 20-qubit quantum simulator},\ }\href
  {https://arxiv.org/abs/2208.04862} {\bibfield  {journal} {\bibinfo  {journal}
  {arXiv preprint arXiv:2208.04862}\ } (\bibinfo {year} {2022})}\BibitemShut
  {NoStop}%
\bibitem [{\citenamefont {Zhu}\ \emph {et~al.}(2022)\citenamefont {Zhu},
  \citenamefont {Wu}, \citenamefont {Bai}, \citenamefont {Wang}, \citenamefont
  {Wang},\ and\ \citenamefont {Chiribella}}]{zhu2022flexible}%
  \BibitemOpen
  \bibfield  {author} {\bibinfo {author} {\bibfnamefont {Y.}~\bibnamefont
  {Zhu}}, \bibinfo {author} {\bibfnamefont {Y.-D.}\ \bibnamefont {Wu}},
  \bibinfo {author} {\bibfnamefont {G.}~\bibnamefont {Bai}}, \bibinfo {author}
  {\bibfnamefont {D.-S.}\ \bibnamefont {Wang}}, \bibinfo {author}
  {\bibfnamefont {Y.}~\bibnamefont {Wang}},\ and\ \bibinfo {author}
  {\bibfnamefont {G.}~\bibnamefont {Chiribella}},\ }\bibfield  {title}
  {\bibinfo {title} {Flexible learning of quantum states with generative query
  neural networks},\ }\href@noop {} {\bibfield  {journal} {\bibinfo  {journal}
  {Nature Communications}\ }\textbf {\bibinfo {volume} {13}},\ \bibinfo {pages}
  {1} (\bibinfo {year} {2022})}\BibitemShut {NoStop}%
\bibitem [{\citenamefont {Lvovsky}(2004)}]{lvovsky2004iterative}%
  \BibitemOpen
  \bibfield  {author} {\bibinfo {author} {\bibfnamefont {A.~I.}\ \bibnamefont
  {Lvovsky}},\ }\bibfield  {title} {\bibinfo {title} {Iterative
  maximum-likelihood reconstruction in quantum homodyne tomography},\ }\href
  {https://iopscience.iop.org/article/10.1088/1464-4266/6/6/014/meta}
  {\bibfield  {journal} {\bibinfo  {journal} {Journal of Optics B: Quantum and
  Semiclassical Optics}\ }\textbf {\bibinfo {volume} {6}},\ \bibinfo {pages}
  {S556} (\bibinfo {year} {2004})}\BibitemShut {NoStop}%
\bibitem [{\citenamefont {Leonhardt}(1997)}]{leonhardt1997measuring}%
  \BibitemOpen
  \bibfield  {author} {\bibinfo {author} {\bibfnamefont {U.}~\bibnamefont
  {Leonhardt}},\ }\href
  {https://www.cambridge.org/ch/academic/subjects/physics/optics-optoelectronics-and-photonics/measuring-quantum-state-light?format=HB&isbn=9780521497305}
  {\emph {\bibinfo {title} {Measuring the quantum state of light}}},\
  Vol.~\bibinfo {volume} {22}\ (\bibinfo  {publisher} {Cambridge university
  press},\ \bibinfo {year} {1997})\BibitemShut {NoStop}%
\bibitem [{\citenamefont {Cybenko}(1989)}]{cybenko1989approximation}%
  \BibitemOpen
  \bibfield  {author} {\bibinfo {author} {\bibfnamefont {G.}~\bibnamefont
  {Cybenko}},\ }\bibfield  {title} {\bibinfo {title} {Approximation by
  superpositions of a sigmoidal function},\ }\href
  {https://link.springer.com/article/10.1007/BF02551274} {\bibfield  {journal}
  {\bibinfo  {journal} {Mathematics of control, signals and systems}\ }\textbf
  {\bibinfo {volume} {2}},\ \bibinfo {pages} {303} (\bibinfo {year}
  {1989})}\BibitemShut {NoStop}%
\bibitem [{\citenamefont {Hornik}(1991)}]{hornik1991approximation}%
  \BibitemOpen
  \bibfield  {author} {\bibinfo {author} {\bibfnamefont {K.}~\bibnamefont
  {Hornik}},\ }\bibfield  {title} {\bibinfo {title} {Approximation capabilities
  of multilayer feedforward networks},\ }\href
  {https://www.sciencedirect.com/science/article/pii/089360809190009T}
  {\bibfield  {journal} {\bibinfo  {journal} {Neural networks}\ }\textbf
  {\bibinfo {volume} {4}},\ \bibinfo {pages} {251} (\bibinfo {year}
  {1991})}\BibitemShut {NoStop}%
\bibitem [{\citenamefont {Ofek}\ \emph {et~al.}(2016)\citenamefont {Ofek},
  \citenamefont {Petrenko}, \citenamefont {Heeres}, \citenamefont {Reinhold},
  \citenamefont {Leghtas}, \citenamefont {Vlastakis}, \citenamefont {Liu},
  \citenamefont {Frunzio}, \citenamefont {Girvin}, \citenamefont {Jiang} \emph
  {et~al.}}]{ofek2016extending}%
  \BibitemOpen
  \bibfield  {author} {\bibinfo {author} {\bibfnamefont {N.}~\bibnamefont
  {Ofek}}, \bibinfo {author} {\bibfnamefont {A.}~\bibnamefont {Petrenko}},
  \bibinfo {author} {\bibfnamefont {R.}~\bibnamefont {Heeres}}, \bibinfo
  {author} {\bibfnamefont {P.}~\bibnamefont {Reinhold}}, \bibinfo {author}
  {\bibfnamefont {Z.}~\bibnamefont {Leghtas}}, \bibinfo {author} {\bibfnamefont
  {B.}~\bibnamefont {Vlastakis}}, \bibinfo {author} {\bibfnamefont
  {Y.}~\bibnamefont {Liu}}, \bibinfo {author} {\bibfnamefont {L.}~\bibnamefont
  {Frunzio}}, \bibinfo {author} {\bibfnamefont {S.}~\bibnamefont {Girvin}},
  \bibinfo {author} {\bibfnamefont {L.}~\bibnamefont {Jiang}}, \emph {et~al.},\
  }\bibfield  {title} {\bibinfo {title} {Extending the lifetime of a quantum
  bit with error correction in superconducting circuits},\ }\href
  {https://www.nature.com/articles/nature18949} {\bibfield  {journal} {\bibinfo
   {journal} {Nature}\ }\textbf {\bibinfo {volume} {536}},\ \bibinfo {pages}
  {441} (\bibinfo {year} {2016})}\BibitemShut {NoStop}%
\bibitem [{\citenamefont {Lvovsky}\ \emph {et~al.}(2020)\citenamefont
  {Lvovsky}, \citenamefont {Grangier}, \citenamefont {Ourjoumtsev},
  \citenamefont {Parigi}, \citenamefont {Sasaki},\ and\ \citenamefont
  {Tualle-Brouri}}]{lvovsky2020production}%
  \BibitemOpen
  \bibfield  {author} {\bibinfo {author} {\bibfnamefont {A.}~\bibnamefont
  {Lvovsky}}, \bibinfo {author} {\bibfnamefont {P.}~\bibnamefont {Grangier}},
  \bibinfo {author} {\bibfnamefont {A.}~\bibnamefont {Ourjoumtsev}}, \bibinfo
  {author} {\bibfnamefont {V.}~\bibnamefont {Parigi}}, \bibinfo {author}
  {\bibfnamefont {M.}~\bibnamefont {Sasaki}},\ and\ \bibinfo {author}
  {\bibfnamefont {R.}~\bibnamefont {Tualle-Brouri}},\ }\bibfield  {title}
  {\bibinfo {title} {Production and applications of non-gaussian quantum states
  of light},\ }\href {https://arxiv.org/abs/2006.16985} {\bibfield  {journal}
  {\bibinfo  {journal} {arXiv preprint arXiv:2006.16985}\ } (\bibinfo {year}
  {2020})}\BibitemShut {NoStop}%
\bibitem [{\citenamefont {Man’ko}\ \emph {et~al.}(1999)\citenamefont
  {Man’ko}, \citenamefont {Moshinsky},\ and\ \citenamefont
  {Sharma}}]{man1999diffraction}%
  \BibitemOpen
  \bibfield  {author} {\bibinfo {author} {\bibfnamefont {V.}~\bibnamefont
  {Man’ko}}, \bibinfo {author} {\bibfnamefont {M.}~\bibnamefont
  {Moshinsky}},\ and\ \bibinfo {author} {\bibfnamefont {A.}~\bibnamefont
  {Sharma}},\ }\bibfield  {title} {\bibinfo {title} {Diffraction in time in
  terms of wigner distributions and tomographic probabilities},\ }\href
  {https://journals.aps.org/pra/abstract/10.1103/PhysRevA.59.1809} {\bibfield
  {journal} {\bibinfo  {journal} {Physical Review A}\ }\textbf {\bibinfo
  {volume} {59}},\ \bibinfo {pages} {1809} (\bibinfo {year}
  {1999})}\BibitemShut {NoStop}%
\bibitem [{\citenamefont {Svozil}\ \emph {et~al.}(1997)\citenamefont {Svozil},
  \citenamefont {Kvasnicka},\ and\ \citenamefont
  {Pospichal}}]{svozil1997introduction}%
  \BibitemOpen
  \bibfield  {author} {\bibinfo {author} {\bibfnamefont {D.}~\bibnamefont
  {Svozil}}, \bibinfo {author} {\bibfnamefont {V.}~\bibnamefont {Kvasnicka}},\
  and\ \bibinfo {author} {\bibfnamefont {J.}~\bibnamefont {Pospichal}},\
  }\bibfield  {title} {\bibinfo {title} {Introduction to multi-layer
  feed-forward neural networks},\ }\href
  {https://www.sciencedirect.com/science/article/pii/S0169743997000610}
  {\bibfield  {journal} {\bibinfo  {journal} {Chemometrics and intelligent
  laboratory systems}\ }\textbf {\bibinfo {volume} {39}},\ \bibinfo {pages}
  {43} (\bibinfo {year} {1997})}\BibitemShut {NoStop}%
\bibitem [{\citenamefont {Higham}(1990)}]{higham1990analysis}%
  \BibitemOpen
  \bibfield  {author} {\bibinfo {author} {\bibfnamefont {N.~J.}\ \bibnamefont
  {Higham}},\ }\bibfield  {title} {\bibinfo {title} {Analysis of the cholesky
  decomposition of a semi-definite matrix},\ }\href
  {http://eprints.maths.manchester.ac.uk/1193/} {\bibfield  {journal} {\bibinfo
   {journal} {University of Manchester eprint}\ } (\bibinfo {year}
  {1990})}\BibitemShut {NoStop}%
\bibitem [{\citenamefont {Blum}(2012)}]{blum2012density}%
  \BibitemOpen
  \bibfield  {author} {\bibinfo {author} {\bibfnamefont {K.}~\bibnamefont
  {Blum}},\ }\href {https://link.springer.com/book/10.1007/978-3-642-20561-3}
  {\emph {\bibinfo {title} {Density matrix theory and applications}}},\
  Vol.~\bibinfo {volume} {64}\ (\bibinfo  {publisher} {Springer Science \&
  Business Media},\ \bibinfo {year} {2012})\BibitemShut {NoStop}%
\bibitem [{\citenamefont {Bimbard}\ \emph {et~al.}(2010)\citenamefont
  {Bimbard}, \citenamefont {Jain}, \citenamefont {MacRae},\ and\ \citenamefont
  {Lvovsky}}]{bimbard2010quantum}%
  \BibitemOpen
  \bibfield  {author} {\bibinfo {author} {\bibfnamefont {E.}~\bibnamefont
  {Bimbard}}, \bibinfo {author} {\bibfnamefont {N.}~\bibnamefont {Jain}},
  \bibinfo {author} {\bibfnamefont {A.}~\bibnamefont {MacRae}},\ and\ \bibinfo
  {author} {\bibfnamefont {A.}~\bibnamefont {Lvovsky}},\ }\bibfield  {title}
  {\bibinfo {title} {Quantum-optical state engineering up to the two-photon
  level},\ }\href {https://www.nature.com/articles/nphoton.2010.6} {\bibfield
  {journal} {\bibinfo  {journal} {Nature Photonics}\ }\textbf {\bibinfo
  {volume} {4}},\ \bibinfo {pages} {243} (\bibinfo {year} {2010})}\BibitemShut
  {NoStop}%
\bibitem [{\citenamefont {Sychev}\ \emph {et~al.}(2017)\citenamefont {Sychev},
  \citenamefont {Ulanov}, \citenamefont {Pushkina}, \citenamefont {Richards},
  \citenamefont {Fedorov},\ and\ \citenamefont
  {Lvovsky}}]{sychev2017enlargement}%
  \BibitemOpen
  \bibfield  {author} {\bibinfo {author} {\bibfnamefont {D.~V.}\ \bibnamefont
  {Sychev}}, \bibinfo {author} {\bibfnamefont {A.~E.}\ \bibnamefont {Ulanov}},
  \bibinfo {author} {\bibfnamefont {A.~A.}\ \bibnamefont {Pushkina}}, \bibinfo
  {author} {\bibfnamefont {M.~W.}\ \bibnamefont {Richards}}, \bibinfo {author}
  {\bibfnamefont {I.~A.}\ \bibnamefont {Fedorov}},\ and\ \bibinfo {author}
  {\bibfnamefont {A.~I.}\ \bibnamefont {Lvovsky}},\ }\bibfield  {title}
  {\bibinfo {title} {Enlargement of optical schr{\"o}dinger's cat states},\
  }\href {https://www.nature.com/articles/nphoton.2017.57} {\bibfield
  {journal} {\bibinfo  {journal} {Nature Photonics}\ }\textbf {\bibinfo
  {volume} {11}},\ \bibinfo {pages} {379} (\bibinfo {year} {2017})}\BibitemShut
  {NoStop}%
\bibitem [{\citenamefont {Gottesman}\ \emph {et~al.}(2001)\citenamefont
  {Gottesman}, \citenamefont {Kitaev},\ and\ \citenamefont
  {Preskill}}]{gottesman2001encoding}%
  \BibitemOpen
  \bibfield  {author} {\bibinfo {author} {\bibfnamefont {D.}~\bibnamefont
  {Gottesman}}, \bibinfo {author} {\bibfnamefont {A.}~\bibnamefont {Kitaev}},\
  and\ \bibinfo {author} {\bibfnamefont {J.}~\bibnamefont {Preskill}},\
  }\bibfield  {title} {\bibinfo {title} {Encoding a qubit in an oscillator},\
  }\href {https://journals.aps.org/pra/abstract/10.1103/PhysRevA.64.012310}
  {\bibfield  {journal} {\bibinfo  {journal} {Physical Review A}\ }\textbf
  {\bibinfo {volume} {64}},\ \bibinfo {pages} {012310} (\bibinfo {year}
  {2001})}\BibitemShut {NoStop}%
\bibitem [{\citenamefont {Glancy}\ and\ \citenamefont
  {Knill}(2006)}]{glancy2006error}%
  \BibitemOpen
  \bibfield  {author} {\bibinfo {author} {\bibfnamefont {S.}~\bibnamefont
  {Glancy}}\ and\ \bibinfo {author} {\bibfnamefont {E.}~\bibnamefont {Knill}},\
  }\bibfield  {title} {\bibinfo {title} {Error analysis for encoding a qubit in
  an oscillator},\ }\href
  {https://journals.aps.org/pra/abstract/10.1103/PhysRevA.73.012325} {\bibfield
   {journal} {\bibinfo  {journal} {Physical Review A}\ }\textbf {\bibinfo
  {volume} {73}},\ \bibinfo {pages} {012325} (\bibinfo {year}
  {2006})}\BibitemShut {NoStop}%
\bibitem [{\citenamefont {Glancy}\ \emph {et~al.}(2009)\citenamefont {Glancy},
  \citenamefont {Vasconcelos}, \citenamefont {Sanz} \emph
  {et~al.}}]{glancy2009generation}%
  \BibitemOpen
  \bibfield  {author} {\bibinfo {author} {\bibfnamefont {S.~C.}\ \bibnamefont
  {Glancy}}, \bibinfo {author} {\bibfnamefont {H.~M.}\ \bibnamefont
  {Vasconcelos}}, \bibinfo {author} {\bibfnamefont {L.}~\bibnamefont {Sanz}},
  \emph {et~al.},\ }\bibfield  {title} {\bibinfo {title} {Generation of gkp
  states with optical states},\ }\href
  {https://tsapps.nist.gov/publication/get_pdf.cfm?pub_id=902362} {\bibfield
  {journal} {\bibinfo  {journal} {nist.gov}\ } (\bibinfo {year}
  {2009})}\BibitemShut {NoStop}%
\bibitem [{\citenamefont {Vlastakis}\ \emph {et~al.}(2013)\citenamefont
  {Vlastakis}, \citenamefont {Kirchmair}, \citenamefont {Leghtas},
  \citenamefont {Nigg}, \citenamefont {Frunzio}, \citenamefont {Girvin},
  \citenamefont {Mirrahimi}, \citenamefont {Devoret},\ and\ \citenamefont
  {Schoelkopf}}]{vlastakis2013deterministically}%
  \BibitemOpen
  \bibfield  {author} {\bibinfo {author} {\bibfnamefont {B.}~\bibnamefont
  {Vlastakis}}, \bibinfo {author} {\bibfnamefont {G.}~\bibnamefont
  {Kirchmair}}, \bibinfo {author} {\bibfnamefont {Z.}~\bibnamefont {Leghtas}},
  \bibinfo {author} {\bibfnamefont {S.~E.}\ \bibnamefont {Nigg}}, \bibinfo
  {author} {\bibfnamefont {L.}~\bibnamefont {Frunzio}}, \bibinfo {author}
  {\bibfnamefont {S.~M.}\ \bibnamefont {Girvin}}, \bibinfo {author}
  {\bibfnamefont {M.}~\bibnamefont {Mirrahimi}}, \bibinfo {author}
  {\bibfnamefont {M.~H.}\ \bibnamefont {Devoret}},\ and\ \bibinfo {author}
  {\bibfnamefont {R.~J.}\ \bibnamefont {Schoelkopf}},\ }\bibfield  {title}
  {\bibinfo {title} {Deterministically encoding quantum information using
  100-photon schr{\"o}dinger cat states},\ }\href@noop {} {\bibfield  {journal}
  {\bibinfo  {journal} {Science}\ }\textbf {\bibinfo {volume} {342}},\ \bibinfo
  {pages} {607} (\bibinfo {year} {2013})}\BibitemShut {NoStop}%
\bibitem [{\citenamefont {Mancini}\ \emph {et~al.}(1996)\citenamefont
  {Mancini}, \citenamefont {Man'Ko},\ and\ \citenamefont
  {Tombesi}}]{mancini1996symplectic}%
  \BibitemOpen
  \bibfield  {author} {\bibinfo {author} {\bibfnamefont {S.}~\bibnamefont
  {Mancini}}, \bibinfo {author} {\bibfnamefont {V.}~\bibnamefont {Man'Ko}},\
  and\ \bibinfo {author} {\bibfnamefont {P.}~\bibnamefont {Tombesi}},\
  }\bibfield  {title} {\bibinfo {title} {Symplectic tomography as classical
  approach to quantum systems},\ }\href
  {https://www.sciencedirect.com/science/article/pii/0375960196001077}
  {\bibfield  {journal} {\bibinfo  {journal} {Physics Letters A}\ }\textbf
  {\bibinfo {volume} {213}},\ \bibinfo {pages} {1} (\bibinfo {year}
  {1996})}\BibitemShut {NoStop}%
\bibitem [{\citenamefont {Man’ko}\ \emph {et~al.}(1998)\citenamefont
  {Man’ko}, \citenamefont {Rosa},\ and\ \citenamefont
  {Vitale}}]{man1998time}%
  \BibitemOpen
  \bibfield  {author} {\bibinfo {author} {\bibfnamefont {V.}~\bibnamefont
  {Man’ko}}, \bibinfo {author} {\bibfnamefont {L.}~\bibnamefont {Rosa}},\
  and\ \bibinfo {author} {\bibfnamefont {P.}~\bibnamefont {Vitale}},\
  }\bibfield  {title} {\bibinfo {title} {Time-dependent invariants and green
  functions in the probability representation of quantum mechanics},\ }\href
  {https://journals.aps.org/pra/abstract/10.1103/PhysRevA.57.3291} {\bibfield
  {journal} {\bibinfo  {journal} {Physical Review A}\ }\textbf {\bibinfo
  {volume} {57}},\ \bibinfo {pages} {3291} (\bibinfo {year}
  {1998})}\BibitemShut {NoStop}%
\end{thebibliography}%

\end{document}